\documentclass[12pt,thmsa]{article}

\input{tcilatex}

\begin{document}

\author{Emilio Santos \\
Departamento de F\'{i}sica. Universidad de Cantabria. \\
Santander. 39005 Spain}
\title{Quantum vacuum fluctuations and dark energy }
\date{September, 21, 2009 }
\maketitle

\begin{abstract}
It is shown that the curvature of space-time induced by vacuum fluctuations
of quantum fields should be proportional to the square of Newton constant $G$%
. This offers a possible explanation for the success of the formula $\rho
\sim $ $Gm^{6}c^{2}h^{-4},$ $\rho $ being the dark energy density and $m$ a
typical mass of elementary particles.


\end{abstract}

The observed accelerated expansion of the universe\cite{Sahni} is currently
assumed to derive from a positive mass density and a negative pressure,
constant throughout space and time, which is known as ``dark energy''. The
mass density, $\rho _{DE},$ and the presure, $p_{DE},$ are \cite{WMAP} 
\begin{equation}
\rho _{DE}\simeq -p_{DE}\simeq 10^{-26}\text{ kg/m}^{3}.  \label{1}
\end{equation}
Many proposals have been made for the origin of dark energy (for a review
see \cite{Copeland} ). The most popular is to identify it with the
cosmological constant $\Lambda $ introduced by Einstein in 1917 or, what is
equivalent in practice, to assume that it derives from the quantum vacuum%
\cite{Jacobson}. Indeed the equality $\rho _{DE}=-p_{DE}$ is appropriate for
the vacuum (in Minkowski space, or when the space-time curvature is small)
because it is invariant under Lorentz transformations.

A problem appears however when one attempts to estimate the value of $\rho
_{DE}.$ In particular if the dark energy is due to the interplay between
quantum mechanics and gravity, it should be either strictly zero or of order
Planck\'{}s density, that is 
\begin{equation}
\rho _{DE}\sim \frac{c^{5}}{G^{2}
\rlap{\protect\rule[1.1ex]{.325em}{.1ex}}h%
}\simeq 10^{97}\text{ kg/m}^{3}.  \label{2}
\end{equation}
This is about 123 orders of magnitude larger than eq.$\left( \ref{1}\right) $%
\cite{SW}. Some explanations have been proposed for the huge value of the
ratio between the Planck density, eq.$\left( \ref{2}\right) ,$ and the dark
energy density, eq.$\left( \ref{1}\right) $. For instance Novello\cite
{Novello} starts proving that, if $\Lambda $ is a fundamental parameter, to
be put on the left side of the Einstein equation (eq.$\left( \ref{a1}\right) 
$ below), the space-time in the absence of matter is de-Sitter and the mass
of the graviton is not zero. Then the ratio between eq.$\left( \ref{2}%
\right) $ and eq.$\left( \ref{1}\right) $ may correspond to the number of
gravitons in the visible universe. On the other hand if $\Lambda $ has to do
with the matter content of the vacuum, it should appear on the right side of
the Einstein equation. In this case $\rho _{DE}$ should be related to other
fundamental constants. Support to the latter assumption is the fact that a
fairly good fit to eq.$\left( \ref{1}\right) $ may be obtained from a
specific combination of the fundamental constants $G, 
\rlap{\protect\rule[1.1ex]{.325em}{.1ex}}h%
$ and $c$ plus a mass parameter $m$ of the order of a typical mass of
fundamental particles, that is\cite{Zel}

\begin{equation}
\rho _{DE}\sim G\frac{m^{6}c^{2}}{
\rlap{\protect\rule[1.1ex]{.325em}{.1ex}}h%
^{4}}.  \label{40}
\end{equation}
Actually the observed value, eq.$\left( \ref{1}\right) ,$ is obtained if the
mass $m$ is 
\begin{equation}
m\sim 7.6\times 10^{-29}\text{ kg}  \label{2a}
\end{equation}
which is about 1/20 times the proton mass or about 80 times the electron
mass.

I believe that dark energy is indeed associated to the matter content of the
quantum vacuum and the purpose of the present paper is to provide a possible
explanation for the estimate eq.$\left( \ref{40}\right) .$ Furthermore I
propose that dark energy derives from the quantum vacuum fluctuations, an
assumption considered recently by Padmanabhan\cite{Padma}, but my approach
here is different. It rests upon the interplay between quantum mechanics and
general relativity. Thus we shall work within quantized gravity. If we
choose an appropriate coordinate system in the universe, with coordinates 
\[
x\equiv \left\{ x_{1},x_{2},x_{3},x_{4}\right\} , 
\]
our aim is to relate the quantum state, $\mid \Phi \rangle ,$ and the
stress-energy tensor operator, $\widehat{T}_{\mu }^{\nu }\left( x\right) ,$
of quantum fields with the space-time structure. That structure is given by
the metric tensor operator, $\widehat{g}_{\mu \nu }.$ The study of the
quantum fields in curved space-times and the gravitational back reaction of
the fields is a difficult subject\cite{Wald} . However for the purposes of
the present paper it is enough to assume that the expectation value of the
stress-energy tensor may be written in the form 
\begin{equation}
\left\langle \Phi \left| \widehat{T}_{\mu }^{\nu }\right| \Phi \right\rangle
=(T^{mat})_{\mu }^{\nu }+(T^{vac})_{\mu }^{\nu },  \label{ad}
\end{equation}
where $(T^{mat})_{\mu }^{\nu }$ is the stress-energy tensor associated to
matter, baryonic and possibly dark, plus radiation, and $(T^{vac})_{\mu
}^{\nu }$ the tensor due to the vacuum. Both terms may be treated as
classical. The latter contribution might be relevant when the space-time
metric departs substantially from Minkowski\'{}s. This may be the case in
the early universe (so explaining ``inflation'') or near galaxies (so
explaining ``dark matter''). Both these possibilities have been explored
recently using f(R)-gravity\cite{Odintsov} , \cite{Faraoni} .

It is convenient to define a stress-energy tensor operator, $(\widehat{T}%
^{fluct})_{\mu }^{\nu },$ such that eq.$\left( \ref{ad}\right) $ may be
rewritten

\begin{equation}
\left\langle vac\left| (\widehat{T}^{fluct})_{\mu }^{\nu }\right|
vac\right\rangle =0,\;(\widehat{T}^{fluct})_{\mu }^{\nu }=\widehat{T}_{\mu
}^{\nu }-\widehat{I}(T^{vac})_{\mu }^{\nu },  \label{a2}
\end{equation}
where $\widehat{I}$ is the identity operator. The existence of vacuum
fluctuations means that, although the expectation of $(\widehat{T}%
^{fluct})_{\mu }^{\nu }$ is zero, there are correlated fluctuations, that is 
\begin{equation}
\left\langle vac\left| (\widehat{T}^{fluct}\left( x\right) )_{\mu }^{\nu
}\times (\widehat{T}^{fluct}\left( y\right) )_{\lambda }^{\sigma }\right|
vac\right\rangle \neq 0\text{ in general.}  \label{aa2}
\end{equation}
For the sake of simplicity I shall ignore the matter terms on the right hand
side of eq.$\left( \ref{ad}\right) $ and write $\widehat{T}_{\mu }^{\nu }$
for $(\widehat{T}^{fluct})_{\mu }^{\nu }$ in the following.

After that I return to the problem of the possible explanation for the
agreement of eq.$\left( \ref{40}\right) $ with observations, eq.$\left( \ref
{1}\right) $. I shall start recalling a well known prediction of quantum
mechanics, namely that correlations between quantum fluctuations may produce
observable effects. An illustrative example is the (van der Waals)
interaction between two molecules at a distance, $d$, much bigger than the
typical size of a molecule. If both molecules possess a permanent electric
dipole moment, then at low enough temperature they are oriented so that the
molecules attract each other. In fact there is a dipole-dipole (negative)
interaction energy which scales as $d^{-3}.$ Now let us consider two neutral
molecules which do not possess permanent dipole moment. In this case
classical physics predicts that there is no electrostatic interaction
between them. Quantum theory however predicts an interaction due to the fact
that quantum fluctuations give rise to instantaneous dipole moments in both
molecules, which are correlated so that the energy is a minimum. This gives
rise to an interaction which scales precisely as the square of the coupling
parameter above mentioned, that is $\left( d^{-3}\right) ^{2}=d^{-6}.$
Indeed it is well known that the interaction energy between nonpolar
molecules decreases with the six power of the distance (when retardation
effects are negligible).

The general behaviour may be understood via perturbation theory. To first
order an average of the quantum fluctuations appears, which is zero (this is
similar to eq.$\left( \ref{a2}\right) $.) To second order however the
perturbation involves the product of two correlated fluctuations, which is
not zero (this is similar to eq.$\left( \ref{aa2}\right) $) and it gives the
interaction to lowest order. The square of the quantum fluctuations involves
the coupling constant squared.

I propose that a similar phenomenon should appear in gravity. Classical
general relativity predicts that any stress-energy, $T_{\mu \nu },$ produces
curvature of the space-time measured by the Ricci tensor, $R_{\mu }^{\nu },$
the scaling parameter being Newton constant, $G$. Indeed this is the
essential content of Einstein equation 
\begin{equation}
R_{\mu }^{\nu }-\frac{1}{2}g_{\mu }^{\nu }R=8\pi c^{-2}GT_{\mu }^{\nu }.
\label{a1}
\end{equation}
The point is that, if there existed a classical non-fluctuating
stress-energy, the curvature would scale as $G$. However if there is no
classical stress-energy, in particular if the vacuum expectation of the
stress-energy tensor is zero, then there is no curvature to order $G$, but
there will be an induced curvature to order $G^{2}$.

Gravity theory is more involved than molecular theory because the equations
of the former are non-linear. In addition there is not yet a satisfactory
quantum gravity theory. Consequently a rigorous derivation of the curvature
due to vacuum fluctuations is not possible at this moment. Nevertheless we
may conjecture that \textit{the curvature due to vacuum fluctuations should
be }$G^{2}$\textit{\ times some weighted average of the correlations of the
quantum vacuum stress-energy tensor at two different space-time points}. I
stress that here I consider fluctuations of all quantum fields (e.g.
electromagnetic radiation field, electron-positron field, etc.) and try to
compute the gravitational effects of these fluctuations. Now the vacuum
expectation, $\left\langle vac\left| \widehat{g}_{\mu \nu }\right|
vac\right\rangle ,$ of the metric tensor, $\widehat{g}_{\mu \nu }$, may be
taken as a classical metric, which should be used to calculate distances and
time intervals. Indeed the accelerated expansion of the universe has been
measured by relating distances and times of supernavae events in distant
galaxies\cite{Sahni}. From the vacuum expectation of the metric tensor we
may derive the Ricci tensor, $R_{\mu }^{\nu },$ which measures the curvature
induced by the quantum vacuum fluctuations. Thus I may write schematically 
\begin{equation}
R_{\mu }^{\nu }\sim G^{2}\times \left\langle quantum\text{ }%
correlation\right\rangle ,  \label{a3}
\end{equation}
where $\left\langle quantum\text{ }correlation\right\rangle $ stands for
some appropriate average of the quantum correlations eq.$\left( \ref{aa2}%
\right) .$

As said above deriving a quantitative expression rather than eq.$\left( \ref
{a3}\right) $ would require a quantum gravity theory not yet available. (In
order to illustrate the main features of the derivation I shall work below a
toy model which is easily soluble). However, a rough estimate of the term $%
\left\langle quantum\text{ }correlation\right\rangle $ may be obtained by
dimensional analysis. In fact, $R_{\mu }^{\nu }$ has dimensions of inverse
length squared. Thus the term $\left\langle quantum\text{ }%
correlation\right\rangle $ should have dimensions of $\left( GL\right)
^{-2}, $ that is $(mass)^{2}(time)^{4}(length)^{-8}.$ This should be
obtained from typical parameters of quantum theory that is the speed of
light, $c$, Planck\'{}s constant, $
\rlap{\protect\rule[1.1ex]{.325em}{.1ex}}h%
$, and some typical mass of elementary particles, $m$. We obtain 
\begin{equation}
\left\langle quantum\text{ }correlation\right\rangle \sim \frac{m^{6}}{%
\rlap{\protect\rule[1.1ex]{.325em}{.1ex}}h%
^{4}}.  \label{a4}
\end{equation}
Here I have excluded Newton constant, $G,$ because at very low curvature, as
in the universe at the present epoch, we should work at the lowest possible
order in $G$, that is just second order as in eq.$\left( \ref{a3}\right) .$

Now we may ask: Is there a classical density, $\rho _{DE},$ which could
mimic the effect ot the vacuum fluctuations as given by eqs.$\left( \ref{a3}%
\right) $ and $\left( \ref{a4}\right) ?.$ A comparison with Einstein eq.$%
\left( \ref{a1}\right) $ gives an affirmative answer provided that 
\begin{equation}
c^{-2}G\rho _{DE}\sim G^{2}\frac{m^{6}}{
\rlap{\protect\rule[1.1ex]{.325em}{.1ex}}h%
^{4}}.  \label{a5}
\end{equation}
This leads precisely to eq.$\left( \ref{40}\right) .$ We conclude that the
quantum vacuum fluctuations should give rise to a curvature of space-time of
the order attributed to the ``dark energy''. Whether there are additional
contributions to dark energy will not be studied in the present paper.

In the following I shall work a simple toy model within quantized gravity
for illustrative pursposes. As quantum gravity theory is not yet available,
I shall introduce a few plausible assumptions. I consider that the metric of
space-time is given by a tensor operator, $\widehat{g}_{\mu \nu },$ and the
quantum vacuum gives rise to a stress-energy tensor operator, $\widehat{T}%
_{\mu }^{\nu }$ , the vacuum expectation of the latter being zero, that is
eq.$\left( \ref{a2}\right) $ holds true. Our aim will be to get the vacuum
expectation of the metric tensor operator in terms of the two-point
correlations 
\[
\left\langle vac\left| \widehat{T}_{\mu }^{\lambda }\left( x\right) _{\text{ 
}}\widehat{T}_{\nu }^{\sigma }\left( y\right) \right| vac\right\rangle . 
\]
In the toy model I assume that quantum vacuum fluctuations possess spherical
symmetry, that is they depend only on the radial coordinate, $r$, and the
time, $t$. Thus we may use standard (or \textit{curvature)} coordinates so
that the (quantized) metric is 
\begin{equation}
d\widehat{s}^{2}=\widehat{A}\left( r,t\right) dr^{2}+\widehat{I}%
r^{2}(d\theta ^{2}+\sin ^{2}\theta d\phi ^{2})-\widehat{B}\left( r,t\right)
dt^{2}.  \label{ds}
\end{equation}
where $\widehat{I}$ is the identity operator. In what follows I shall use
units $c=1$, but write explicitly Newton\'{}s constant, $G$, for the sake of
clarity.

The next task will be to get the operators $\widehat{A}\left( r,t\right) $
and $\widehat{B}\left( r,t\right) $ from the stress-energy tensor operator $%
\widehat{T}_{\mu \lambda }\left( r,t\right) $. As a guide I shall start from
relations valid in classical gravity for a space-time of spherical symmetry
in standard coordinates. It is a fortunate fact that these relations do not
involve explicitly the time coordinate. The relations are\cite{Synge} 
\begin{eqnarray}
A\left( r,t\right)  &=&\left( 1-\frac{2Gm\left( r,t\right) }{r}\right) ^{-1},%
\text{ }m\left( r,t\right) \equiv 4\pi \int_{0}^{r}\rho \left( x,t\right)
x^{2}dx,  \nonumber \\
B(r,t) &=&\exp \left[ 2G\int_{0}^{r}\frac{m\left( x,t\right) +4\pi
x^{3}p\left( x,t\right) }{x^{2}-2Gm\left( x,t\right) x}dx\right] ,
\label{29}
\end{eqnarray}
where $\rho \equiv $ $T_{t}^{t}$ is the mass density and $p\equiv $ $%
-T_{r}^{r}$ the radial pressure. In our model the transverse pressure, $%
q\equiv $ $-T_{\theta }^{\theta }=$ $-T_{\phi }^{\phi }$, may be different
from the radial pressure, but it plays no role in the following. In
quantized gravity the operators $\widehat{\rho }\left( r,t\right) \equiv 
\widehat{T_{t}^{t}}\left( r,t\right) $ and $\widehat{p}\left( r,t\right)
\equiv -\widehat{T_{t}^{t}}\left( r,t\right) $ at different places may not
commute. Therefore getting the quantized counterparts of eqs.$\left( \ref{29}%
\right) $ is not trivial. The problem is easier if we work to second order
in Newton\'{}s constant, $G$, which is enough for the purposes of this
paper. Thus I will write the first eqs.$\left( \ref{29}\right) $ in the form 
\begin{equation}
A\left( r,t\right) =1+\frac{2Gm}{r}+\frac{4G^{2}m^{2}}{r^{2}}+O\left(
G^{3}\right) ,  \label{28}
\end{equation}
Similarly the second eq.$\left( \ref{29}\right) $ may be written 
\begin{eqnarray}
B(r,t) &=&1+2G\int_{0}^{r}\left[ x^{-2}m(x)+4\pi xp\left( x\right) \right] dx
\nonumber \\
&&+4G^{2}\int_{0}^{r}\left[ x^{-3}m(x)^{2}+4\pi m(x)p(x)\right] dx  \nonumber
\\
&&+2G^{2}\left[ \int_{0}^{r}\left[ x^{-2}m(x)+4\pi xp\left( x\right) \right]
dx\right] ^{2}+O\left( G^{3}\right) .  \label{31}
\end{eqnarray}
Now I propose that the quantized counterparts of these equations are
obtained with the replacements $\rho \left( r,t\right) \rightarrow \widehat{%
\rho }\left( r,t\right) ,$ $p\left( r,t\right) \rightarrow \widehat{p}\left(
r,t\right) $ plus the substitution of symmetrically ordered products of
operators for the usual products of the corresponding classical quantities.
Thus for instance 
\[
p\left( r_{1},t_{1}\right) p\left( r_{2},t_{2}\right) \rightarrow \frac{1}{2}%
\left( \widehat{p}\left( r_{1},t_{1}\right) \widehat{p}\left(
r_{2},t_{2}\right) +\widehat{p}\left( r_{2},t_{2}\right) \widehat{p}\left(
r_{1},t_{1}\right) \right) .
\]

After that it is straightforward to get the desired vacuum expectations,
that is

\begin{eqnarray}
\left\langle vac\left| \widehat{A}\left( r,t\right) \right| vac\right\rangle
&\simeq &1+\left\langle vac\left| \frac{2G}{r}\widehat{m}(r)\mathbf{+}\frac{%
4G^{2}}{r^{2}}\widehat{m}(r)^{2}\right| vac\right\rangle ,\text{ }  \nonumber
\\
\widehat{m}(r) &\equiv &\int_{0}^{r}\widehat{\rho }\left( x\right) 4\pi
x^{2}dx,  \label{31b}
\end{eqnarray}
\begin{eqnarray}
\left\langle vac\left| \widehat{B}\left( r,t\right) \right| vac\right\rangle
&\simeq &1+\frac{2G}{r}\left\langle vac\left| \int_{0}^{r}x^{-2}\widehat{m}%
(x)dx+\int_{0}^{r}4\pi x\widehat{p}\left( x\right) dx\right| vac\right\rangle
\nonumber \\
&&+2G^{2}\left\langle vac\left| \left[ \int_{0}^{r}x^{-2}\widehat{m}(x)dx%
\mathbf{+}\int_{0}^{r}4\pi x\widehat{p}\left( x\right) dx\right] ^{2}\right|
vac\right\rangle  \nonumber \\
&&+2G^{2}\left\langle vac\left| \int_{0}^{r}4\pi \left[ \widehat{m}(x)%
\widehat{p}\left( x\right) +\widehat{p}\left( x\right) \widehat{m}(x)\right]
dx\right| vac\right\rangle  \nonumber \\
&&.+2G^{2}\left\langle vac\left| 2\int_{0}^{r}x^{-3}\widehat{m}%
(x)^{2}dx\right| vac\right\rangle .  \label{31c}
\end{eqnarray}
It may be realized that in these expressions the quantum operators $\widehat{%
\rho }$ and $\widehat{p}$ appear always in symmetrical order. The dependence
on time of the operators $\widehat{\rho }$ and $\widehat{p}$ has not been
explicitly shown, and it will not be written in what follows. It is
understood that all quantities are defined at the same coordinate time with
the metric $\left( \ref{ds}\right) $. In eqs.$\left( \ref{31b}\right) $ and $%
\left( \ref{31c}\right) $ the terms of order $G$ cancel, taking eq.$\left( 
\ref{a2}\right) $ into account. It is straightforward to write the terms of
order $G^{2}$ in terms of the correlations of the stress-energy tensor that
is 
\begin{eqnarray}
\frac{1}{2}\left\langle vac\left| \widehat{\rho }(r_{1})\widehat{\rho }%
(r_{2})+\widehat{\rho }(r_{2})\widehat{\rho }(r_{1})\right| vac\right\rangle
&\equiv &C_{\rho \rho }\left( r_{1},r_{2}\right) ,  \nonumber \\
\frac{1}{2}\left\langle vac\left| \widehat{p}(r_{1})\widehat{p}(r_{2})+%
\widehat{p}(r_{2})\widehat{p}(r_{1})\right| vac\right\rangle &\equiv
&C_{pp}\left( r_{1},r_{2}\right) ,  \nonumber \\
\frac{1}{2}\left\langle vac\left| \widehat{\rho }(r_{1})\widehat{p}(r_{2})+%
\widehat{p}(r_{2})\widehat{\rho }(r_{1})\right| vac\right\rangle &\equiv
&C_{\rho p}\left( r_{1},r_{2}\right) ,  \label{a6}
\end{eqnarray}
but I shall not do that explicitly.

In a semiclassical description the vacuum expectations of the metric tensor
operator should be taken as a classical metric tensor. That tensor is the
same which would be produced by some classical stress-energy tensor which
may be obtained by comparison of eq.$\left( \ref{28}\right) $ with eq.$%
\left( \ref{31b}\right) $ and eq.$\left( \ref{31}\right) $ with eq.$\left( 
\ref{31c}\right) $. For instance the former gives 
\[
m_{cl}\left( r\right) =\frac{2G}{r}\left\langle vac\left| \widehat{m}%
(r)^{2}\right| vac\right\rangle . 
\]
Hence, taking into account the relations between $m_{cl}$ and $\rho _{cl}$
and between $\widehat{m}$ and $\widehat{\rho }$, we get 
\[
\rho _{cl}\left( r\right) =8\pi G\left[ \frac{2}{r}\int_{0}^{r}x^{2}C_{\rho
\rho }\left( r,x\right) dx-\frac{1}{r^{4}}\int_{0}^{r}x^{2}dx%
\int_{0}^{r}y^{2}dyC_{\rho \rho }\left( x,y\right) \right] . 
\]
A similar, but more involved, expression may be obtained for $p_{cl}(r)$,
but I shall not write it. The interesting result is that the (ficticious)
classical density, $\rho _{cl},$ is the product of Newton constant $G$ times
some average of the correlation between quantum fluctuations of the density.
This makes plausible our proposals eqs.$\left( \ref{a3}\right) $ to $\left( 
\ref{a5}\right) .$

In summary, the arguments of the present paper strongly suggest that quantum
vacuum fluctuations necessarily give rise to a curvature of space-time which
roughly agrees with that attributed to dark energy. However a rigorous
derivation would be required, which is not possible in the absence of a
quantum gravity theory.

\end{document}